\documentclass[prl,noshowkeys,twocolumn,superscriptaddress]{revtex4}
\usepackage{graphicx}

\def\gap{\;\rlap{\lower 2.5pt
 \hbox{$\sim$}}\raise 1.5pt\hbox{$>$}\;}
\def\lap{\;\rlap{\lower 2.5pt
   \hbox{$\sim$}}\raise 1.5pt\hbox{$<$}\;}
\def\gsim{\;\rlap{\lower 2.5pt
 \hbox{$\sim$}}\raise 1.5pt\hbox{$>$}\;}
\def\lsim{\;\rlap{\lower 2.5pt
   \hbox{$\sim$}}\raise 1.5pt\hbox{$<$}\;}
\def\msun{{\rm\,M_\odot}}

\def\spose#1{\hbox to 0pt{#1\hss}}
\def\lta{\mathrel{\spose{\lower 3pt\hbox{$\mathchar''218$}}
     \raise 2.0pt\hbox{$\mathchar''13C$}}}
\def\gta{\mathrel{\spose{\lower 3pt\hbox{$\mathchar''218$}}
     \raise 2.0pt\hbox{$\mathchar''13E$}}}
\newcommand{\beq}{\begin{equation}}
\newcommand{\eeq}{\end{equation}}
\newcommand{\be}{\begin{equation}}
\newcommand{\ee}{\end{equation}}

\newcommand{\ls}{\mathrel{\raise1.16pt\hbox{$<$}\kern-7.0pt 
\lower3.06pt\hbox{{$\scriptstyle \sim$}}}}         
\newcommand{\gs}{\mathrel{\raise1.16pt\hbox{$>$}\kern-7.0pt 
\lower3.06pt\hbox{{$\scriptstyle \sim$}}}}         

\long\def\comment#1{}

\def\mh{M_{\bullet}}
\def\msun{M_{\odot}}
\def\fun#1#2{\lower3.6pt\vbox{\baselineskip0pt\lineskip.9pt
  \ialign{$\mathsurround=0pt#1\hfil##\hfil$\crcr#2\crcr\sim\crcr}}}
\def\lap{\mathrel{\mathpalette\fun <}}
\def\gap{\mathrel{\mathpalette\fun >}}
\newcommand{\ba}{\begin{eqnarray}}
\newcommand{\ea}{\end{eqnarray}}

\begin{document}
\bibliographystyle{apsrev.bst}
\title{Dark Matter Spikes and Annihilation Radiation from the Galactic Center}
\author{David Merritt}
\email{merritt@physics.rutgers.edu}
\affiliation{Dept. of Physics \& Astronomy, Rutgers University, 
New Brunswick, NJ}
\author{Milos Milosavljevi\'c}
\email{milos@physics.rutgers.edu}
\affiliation{Dept. of Physics \& Astronomy, Rutgers University, 
New Brunswick, NJ}
\author{Licia Verde}
\email{lverde@astro.princeton.edu}
\affiliation{Dept. of Physics \& Astronomy, Rutgers University, 
New Brunswick, NJ}
\affiliation{Dept. of Astrophysical Sciences, Princeton University,
Princeton, NJ}
\author{Raul Jimenez}
\email{raulj@physics.rutgers.edu}
\affiliation{Dept. of Physics \& Astronomy, Rutgers University, 
New Brunswick, NJ}


\begin{abstract}
The annihilation rate of weakly interacting cold dark matter particles 
at the galactic center could be greatly enhanced by the growth of a density
spike around the central supermassive black hole (SBH). Here we discuss the 
effects of hierarchical mergers on the central spike.
Mergers between halos containing SBHs lead to the formation of SBH binaries 
which transfer energy to the dark matter particles, lowering their density. 
The predicted flux of annihiliation photons from the galactic center is 
several orders of magnitude smaller than in models that ignore the effects 
of SBHs and mergers.  Measurement of the annihilation radiation could in
principle be used to constrain the merger history of the galaxy.
\end{abstract}

\maketitle

The dark matter puzzle is one of the central challenges facing 
particle physics and cosmology \cite{Primack:01}.
A popular candidate for non-baryonic cold dark matter (CDM) is the
lightest supersymmetric particle, plausibly the neutralino $\chi$
\cite{PP:82,Goldberg:83}.
The mass of the neutralino is constrained by accelerator
searches and theoretical considerations of thermal freeze-out
to lie in the range $30 \mathrm{GeV} \lap M_{\chi} \lap 10 \mathrm{TeV}$
\cite{Ellis:83, Jungman:96}.
Neutralinos are generically found to decouple at a temperature
that is roughly $M_{\chi}/20$, which means that they are nonrelativistic
already at decoupling and hence behave like CDM.

Dark matter particles may be detected directly, via laboratory
experiments \cite{Feng:00a}, or indirectly, 
via their annihilation products \cite{Feng:01}.
Indirect schemes are typically based on searches for gamma rays 
from neutralino annihilations in the dark matter halo of the Milky Way (MW)
galaxy \cite{Urban:92,Berezinsky:94,Bergstrom:98}.
Since the photon flux depends on the squared density of neutralinos 
integrated along the line of sight, the signal is
greatly enhanced in directions where the dark matter is clumped.
This includes the galactic center, where the density in a smooth halo
would be maximum, as well as any lines of sight intersecting the
centers of relic halos that orbit as subclumps in the MW halo 
\cite{Bergstrom:99,Calcaneo:00}.
The signal from the galactic center is further enhanced if there
is a CDM ``spike'' associated with the central supermassive black hole (SBH).
Adiabatic growth of a SBH at the center of a pre-existing halo
produces a power-law distribution of matter around the SBH, a ``spike,''
with density $\rho\sim r^{-\gamma}$, $2.2\lap\gamma\lap 2.5$,
$r\lap r_h\equiv G\mh/\sigma^2$ \cite{Gondolo:99}; $\mh$ is the SBH mass
and $\sigma$ is the 1D velocity dispersion of the dark matter particles.
For the MW SBH, $\mh\sim 2-5\times 10^6\msun$
\cite{Ghez:98,Genzel:00,Eckart:02} 
and $r_h\sim 1$ pc.
These spikes lead to predictions of higher-than-observed rates of 
$\gamma$-ray annihilation products\cite{Gandolo:00}.

One element missing from earlier discussions of dark matter spikes
is the destructive effect of hierarchical mergers.
A dark matter halo as massive as that of the MW, $M\sim 10^{12}\msun$, 
has almost certainly
experienced a significant merger event since a redshift of $z\sim 2$.
Furthermore, SBHs with masses of $\sim 10^9\msun$
were present in at least some halos already at redshifts of $5-6$
\cite{Fan:00,Fan:01},
and SBHs probably acquired most of their mass
by a redshift of $2-3$, the epoch of peak quasar activity\cite{MF:01}.
In the CDM paradigm, big halos grow through the buildup of smaller ones
\cite{Press:74};
if more than one of the progenitor halos carried a central SBH,
a binary SBH will form following the merger \cite{Begelman:80}.
Formation and decay of the SBH binary transfers energy to the background
particles and lowers the density of matter in a
region within a few $\times r_h$ around the binary \cite{Merritt:01,Milos:01},
roughly the scale of the spike,
before the SBHs coalesce due to emission of gravitational radiation
\cite{Peters:64}.
In this paper we compute the dark matter density profiles of merging CDM 
halos containing SBHs and show that the net result of including the SBHs 
is to substantially {\it lower} the predicted flux of annhilation photons
compared with halos lacking SBHs.
Our results are relevant also to models that relate the origin
of ultra high energy cosmic rays to the decay of unstable
superheavy relic particles in the galactic halo \cite{Berezinsky:97}.

We first calculated the probability that the MW halo has experienced a major 
merger event by generating multiple realizations of the merger history 
of a halo of $10^{12}\msun$ using the algorithm described by Somerville 
and Kolatt \cite{Somerville:99}, as applied to the standard $\Lambda$CDM 
cosmological model ($\Omega_0=0.3$, $\Lambda_0=0.7$). 
This algorithm accurately reproduces the merger histories of halos 
seen in $N$-body simulations of structure formation \cite{Somerville:00}
although at a value of the scale factor that may differ as much as $20\%$ from
the true value \cite{Wechsler:02}.
We recorded all merger events such that the mass of the smaller of the 
subclumps involved in the merger was above a given limit $M_{\rm lim}$.
From $600$ of these realizations, we then computed the probability that a halo 
of $10^{12}\msun$ at redshift $z=0$ had in its merger history at least 
one such event since a redshift of $z=2$.
Fig. 1 shows this probability as a function of $M_{\rm lim}$. 
With $68\%$ confidence, a halo the size of the MW has experienced at least 
one merger since $z=2$ in which the mass of the smaller of the merging
halos was above $2\times 10^{11}\msun$, implying a mass ratio between
subclumps of $4:1$ or less.
Mergers with mass ratios of $9:1$ or less occurred with
greater than $90\%$ probability.

\begin{table}
\caption{\label{Table1}Parameters of the $N$-body integrations.
$M_1$ ($M_2$) is the mass of the large (small) halo; $\rho$ is the
central halo density before growth of the SBH; 
$\mh$ is the mass of the SBH particle; $N$ is
the combined number of DM particles in both halos; 
$a_{\rm final}$ is the final separation of the BH binary.}
\begin{ruledtabular}
\begin{tabular}{cccccc}
Run & $M_1/M_2$ & $\rho_1/\rho_2$ & $\mh/M$ & $N$ &
$a_{\rm final} ({\rm pc})$ 
\\ 
A & 1  & 1   & $0.01$ & $10^5$           & $0.91$ \\
B & 1  & 1   & $0.03$ & $2\times 10^5$   & $0.71$ \\
C & 3  & 1   & $0.03$ & $4\times 10^4$   & $0.31$ \\
D & 3  & 1/3 & $0.03$ & $4\times 10^4$   & $0.42$ \\
E & 5  & 1   & $0.03$ & $6\times 10^4$   & $0.36$ \\
F & 10 & 1   & $0.03$ & $1.1\times 10^5$ & $0.90$ \\
\end{tabular}
\end{ruledtabular}
\end{table}

To investigate the detailed effects of mergers on dark matter spikes,
we used an $N$-body code to simulate interactions between CDM halos 
containing central SBHs.
$N$-body models were generated from the spherical density law
\begin{equation}
\rho_{\rm DM}(x) = \rho_0x^{-1.5}(1+x^{1.5})^{-1},\ \ \ \ x\equiv r/r_{\rm DM},
\end{equation}
the ``Moore profile'' \cite{Moore:99}, one of a set of profiles found to 
accurately describe the halos generated in CDM structure formation 
simulations; $r_{\rm DM}$ defines the radius of transition between the inner 
cusp and the steeper outer falloff.
Our choice of the Moore profile is conservative in the sense that
other proposed fitting functions \cite{Navarro:97,Klypin:01a}
have shallower central density cusps implying less prominent spikes.
The Moore profile has a divergent mass; we truncated it spatially 
and generated velocities of the dark matter particles 
from a distribution function \cite{Merritt:85} which generates an  
isotropic velocity distribution near the center and increasingly 
circular orbits approaching the truncation radius.
This procedure guarantees a state of detailed dynamical equilibrium
in spite of the model's hard edge.
Since the dependence of the central densities of CDM halos on halo mass is
not well determined, we investigated two values of the central density ratio
$\rho_1/\rho_2=\{1,1/3\}$ for one of the unequal-mass mergers,
where $\rho$ is defined as the asymptotic central density of the
halos before growth of the SBH.
A ``black hole'' was inserted into each $N$-body halo by slowly
increasing the mass of a (Newtonian) central point particle
from $0$ to $\mh$ (see Ref. \cite{Merritt:98}).
We verified that the resulting density profile satisfied
$\rho_{\rm DM}\sim r^{-2.4}$, $r\lap r_h$ as predicted by adiabatic
theory (\cite{Gondolo:99}; Fig. 2).
Since the scale length of the MW halo is believed to be 
$r_{\rm DM}\sim 20$ kpc \cite{Klypin:01b}, the dark matter density
profile is essentially scale-free, 
$\rho_{\rm DM}\sim r^{-1.5}$, at all radii of interest
and the mass chosen for the particle representing the MW SBH
is fairly arbitrary.
We tried two values, $\mh/M_{\rm DM}=\{0.01,0.03\}$; the larger value
was preferred since it resulted in a lower amplitude of
Brownian motion of the SBH particle \cite{Dorband:01}.
The same mass ratio was adopted for the SBH in the smaller halo.
The parameters of the $N$-body runs are given in Table \ref{Table1}.

\begin{figure}
\includegraphics[height=8cm]{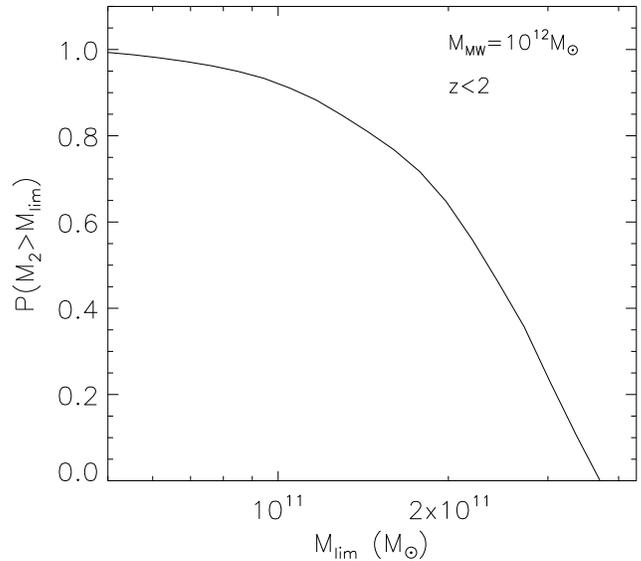}
\caption{Probability that the merger tree of a CDM halo of mass 
$10^{12}\msun$ at $z=0$ contains at least one merger event between
$z=0$ and $z=2$ in which the mass $M_2$ of the smaller subclump exceeded 
$M_{\rm lim}$.
}
\label{fig1}
\end{figure}

The mass scaling of our models was fixed by the measured mass of the 
MW SBH \cite{Ghez:98,Genzel:00,Eckart:02};
we assumed $\mh=3\times 10^6\msun$.
The length scale was determined by assuming that the dark matter in the MW
produces a circular rotation velocity of $90$ km s$^{-1}$ at the solar circle,
$R_{\odot}=8$ kpc \cite{Dehnen:98} and that its density profile is
$\rho_{\rm DM}\propto r^{-1.5}$ inward of $R_{\odot}$.

Simulated mergers of the halo models were carried out using a new, 
general-purpose $N$-body algorithm \cite{Milos:02a} that combines the 
elements of a hashed tree code with a quadrupolar expansion of force moments 
for the bulk dynamics, and the highly accurate Hermite predictor-corrector 
scheme for near-neighbor and massive-particle interactions.  The code 
implements individual block time steps, individual stepping and softening 
criteria, and full functional parallelization. Interactions between SBH
and dark-matter particles were unsoftened.
Calculations were carried out using 16 processors on the Rutgers Sun HPC-10000 
computer. Integrations were terminated when the separation between the SBH 
particles was less than $1$ pc (see Table 1).

Fig. 2 shows the effect of mergers on the dark-matter density
profiles.
The steep, $\rho\sim r^{-2.4}$ density spikes are destroyed in each
case by the transfer of kinetic energy from the binary SBH to 
the dark matter particles.
(Mergers without central SBHs tend to preserve spike slopes
\cite{Barnes:99,Milos:01}.)
The energy transfer takes place in two stages \cite{Milos:01}:
before the two SBHs form a binary system, dynamical friction
acting on the individual SBHs causes their orbits to decay;
and after formation of a bound pair, dark matter particles which pass
the binary within a few times the binary's semi-major axis are ejected
by the gravitational slingshot.
The result is a lowered density out to a ``core'' radius $r_c$ of 
$\sim 10-100$ pc and a density profile that rises inward of $r_c$ 
as a weak power-law, $\rho\sim r^{-0.5}$.
The amount of damage done to the pre-existing spike increases with the mass 
of the secondary SBH, 
consistent with the expectation that the mass ejected by the binary
is of order a few times $M_2$ \cite{Quinlan:96}.
Final mean dark-matter densities within $100$ pc are $5-10\msun$ pc$^{-3}$
for all of the runs, rising inward to
$\rho \sim 10^2\msun$ pc$^{-3}$ for the $1:1$ mergers and 
$\sim 10^3\msun$ pc$^{-3}$ for the $10:1$ merger. 

\begin{figure}
\includegraphics[width=8cm]{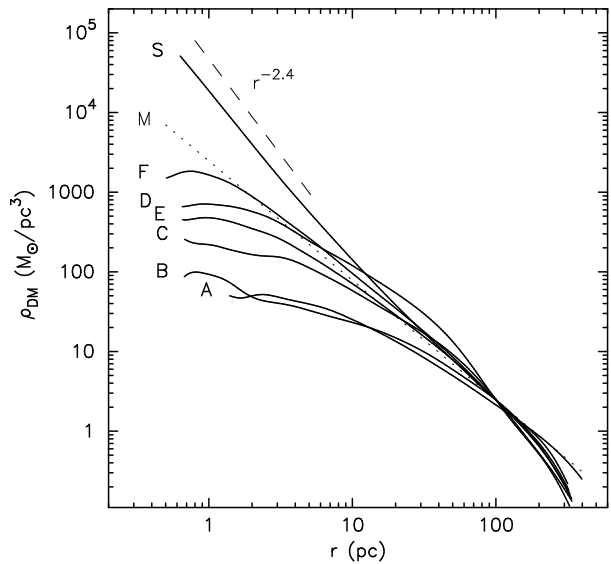}
\caption{Density profiles of the merged dark matter halos.
The origin is defined as the center of mass of the binary
SBH; spherical symmetry is assumed.
The curves labelled $M$ and $S$ are the density profiles of the
large halo before and after growth of the SBH. 
}
\label{fig2}
\end{figure}

The differential photon flux along a direction that makes
an angle $\psi$ with respect to the galactic center is
\begin{equation}
{d\Phi_{\gamma}\over d\Omega} = \sum_i N^i_{\gamma} 
{\sigma_iv\over 4\pi M_{\chi}^2}\int_{\psi} \rho^2 dl,
\end{equation}
where $\Omega$ is the solid angle, 
$\rho$ is the neutralino density, and 
$\sigma v$ is the annihilation cross section
(independent of $v$ for nonrelativistic particles);
the sum is over all annihilation channels.
$N_{\gamma}=2$ for $\chi\chi\rightarrow\gamma\gamma$ and 
$N_{\gamma}=1$ for $\chi\chi\rightarrow Z\gamma$
\cite{Bergstrom:98}.
We are principally concerned with the final line-of-sight integral,
$J(\psi)=\int_{\psi}\rho^2 dl$, which contains
all of the information about the halo density profile.
Following earlier authors \cite{Bergstrom:98,Feng:01},
we write $J$ in dimensionless form as
\begin{equation}
J(\psi) = {1\over 8.5 \mathrm{kpc}}\left({1\over 0.3 \mathrm{GeV/cm}^3}\right)^2 \int_{\psi}\rho^2 dl
\end{equation}
where the normalizing factors for length and density are roughly
the radius of the solar circle and the local density of dark
matter respectively.
Finally we average the flux over the field of view assuming a 
circular aperture of size $\Delta\Omega$ centered at $\psi=0$:
\begin{equation}
\langle J\rangle \equiv {1\over \Delta\Omega} \int_{\Delta\Omega}J(\psi)\ 
d\psi.
\end{equation}
Fig. 3 plots $\langle J\rangle$ as a function of $\Delta\Omega$ 
at the end of the $N$-body simulations.
For a typical atmospheric Cherenkov telescope angular acceptance
of $\Delta\Omega=10^{-3}$ sr, $\langle J\rangle$ ranges from 
$\sim 10^{3.2}$ for the $1:1$ merger, to $\sim 10^{3.9}$ for the 
$3:1$ mergers, compared with $\sim 10^{4.2}$ for the initial
halo model without a SBH-induced spike
(computed assuming an inner cutoff at the Schwarzschild radius of the SBH).
Thus the addition of SBHs to CDM halos
results in a net {\it decrease} in the annihilation flux compared
with SBH-free models, if mergers are taken into account.
The values of $\langle J\rangle$ in Fig. 3 are nevertheless large
enough to allow testing of large parts of the neutralino parameter 
space using instruments like GLAST \cite{Ullio:01a}.

We note that the predicted flux is a strong function of the merger 
parameters when observed with $\Delta\Omega\lap 10^{-6}$ (Fig. 3), 
corresponding roughly to the sphere of influence of the MW SBH.
This fact might allow the merger history of the MW to be inferred
from measurements of the annihilation flux on different angular scales.

\begin{figure}
\includegraphics[width=8cm]{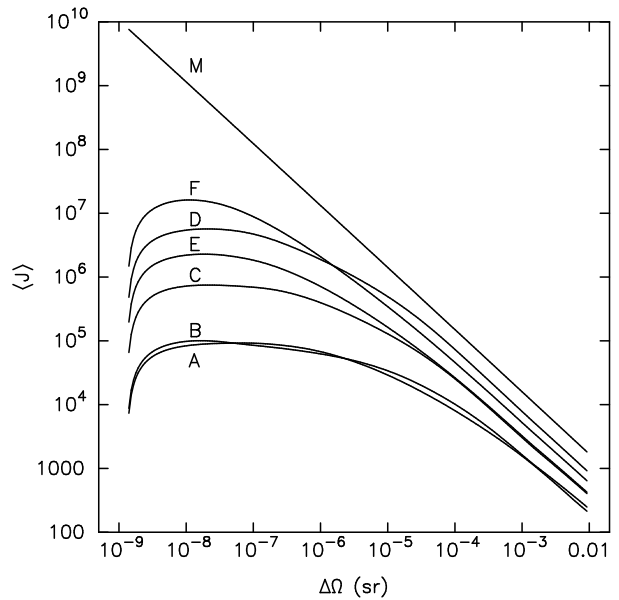}
\caption{Dimensionless integrated flux of the merger models 
as a function of the angular
acceptance of the detector $\Delta\Omega$ (Eq. 4).
The curve labelled $M$ is the flux predicted by a Moore-profile
halo without a SBH-induced spike (Fig. 2).
}
\label{fig3}
\end{figure}

The addition of SBHs to CDM halos could result in even lower densities
than shown in Fig. 2.
Halos as massive as that of the MW are believed
to have formed through a succession of mergers, and the
damage done by binary SBHs would be to some extent
cumulative, resulting in shallower central profiles
than found here \cite{Milos:01,Milos:02b}.
In addition, binary SBHs may eventually coalesce due to emission of 
gravitational radiation \cite{Peters:64}. The radiation so emitted carries 
linear momentum leading to a recoil of the SBH at a velocity
$v_{\rm recoil}\sim 10^2-10^3$ km/s 
\cite{Bekenstein:73,Fitchett:83,Eardley:83},
and possibly even higher if the SBHs were rapidly spinning prior to coalescence
\cite{Redmount:89}.
Recoil velocities of this order would eject the coalesced SBH from the
nucleus and its subsequent infall would displace dark matter particles
(e.g. \cite{Nakano:99,Ullio:01b}).
Quantitative evaluation of this effect will require more accurate 
estimates of $v_{\rm recoil}$ based on fully general-relativistic 
calculations of black hole mergers.

We thank P. Ullio for useful discussions.
D. M. was supported by NSF grant AST 00-71099 and NASA grants
NAG5-6037 and NAG5-9046, and L. V. was supported in part by NASA 
grant NAG5-7154.
We thank the Center for Advanced Information Processing at Rutgers 
University for their generous allocation of CPU time.


\begin{thebibliography}{99}

\bibitem{Primack:01}
	J. R. Primack, 
	in Proceedings of the International School of Space Science 2001, 
	ed. A. Morselli (Frascati Physics Series)
	p. 449 (2002).

\bibitem{PP:82} 
	H. Pagels and J. R. Primack, Phys. Rev. Lett. {\bf 48}, 223 (1982)

\bibitem{Goldberg:83}
	H. Goldberg, Phys Rev. Lett. {\bf 50}, 1419 (1983).

\bibitem{Ellis:83}
	J. Ellis, J. S. Hagelin, D. V. Nanopoulos, and M. Srednicki,
	Phys. Lett. {\bf 127B}, 233 (1983).

\bibitem{Jungman:96}
	G. Jungman, M. Kamionkowski, and K. Griest,
	Phys. Rep. {\bf 267}, 195 (1996).

\bibitem{Feng:00a} 
	J. L. Feng, K. T. Matchev, and F. Wilczek, 
	Phys. Lett. B {\bf 482}, 388 (2000).

\bibitem{Feng:01} 
	J. L. Feng, K. T. Matchev, and F. Wilczek,
	Phys. Rev. D. {\bf 63}, 045024 (2001).

\bibitem{Urban:92}
	M. Urban, A. Bouquet, B. Degrange, P. Fleury, J. Kaplan, 
	A. L. Melchior, and E. Par\'e,
	Phys. Lett. B {\bf 293}, 149 (1992).

\bibitem{Berezinsky:94}
	V. Berezinsky, A. Bottino, and G. Mignola,
	Phys. Lett. B {\bf 325}, 136 (1994).

\bibitem{Bergstrom:98}
	L. Bergstr\"om, P. Ullio, and J. H. Buckley,
	Astroparticle Phys. {\bf 9}, 137 (1998).

\bibitem{Bergstrom:99}
	L. Bergstr\"om, J. Edsj\"o, P. Gondolo, and P. Ullio,
	Phys. Rev. D {\bf 59}, 043506 (1999).

\bibitem{Calcaneo:00}
	C. Calc\'aneo-Rold\'an and B. Moore,
	Phys. Rev. D {\bf 62}, 123005 (2000).

\bibitem{Gondolo:99}
	P. Gondolo and J. Silk,
	Phys. Rev. Lett. {\bf 83}, 1719 (1999).

\bibitem{Ghez:98}
	A. M. Ghez, B. L. Klein, M. Morris, and E. E. Becklin,
	Astrophys. J. {\bf 509}, 678 (1998).

\bibitem{Genzel:00}
	R. Genzel, C. Pichon, A. Eckart, O. E. Gerhard, and T. Ott,
	Mon. Not. Royal Astron. Soc. {\bf 317}, 348 (2000).

\bibitem{Eckart:02}
	A. Eckart, R. Genzel, T. Ott, and R. Schoedel,
	astro-ph/0201031

\bibitem{Gandolo:00}
	P. Gandolo,
	hep-ph/0002226

\bibitem{Fan:00}
	X. Fan {\it et al.},
	Astron. J. {\bf 120}, 1167 (2000).

\bibitem{Fan:01}
	X. Fan {\it et al.},
	Astron. J., {\bf 122}, 2833 (2001)

\bibitem{MF:01} 
	D. Merritt and L. Ferrarese, in
	ASP Conf. Ser. Vol. 249, ``The Central Kiloparsec of Starbursts
	and AGN: The La Palma Connection'', ed. J. H. Knapen, J. E. Beckman,
	I. Shlosman, and T. J. Mahoney (Astronomical Society of the Pacific),
	p. 335 (2001).

\bibitem{Press:74}
	W. H. Press and P. Schechter,
	Astrophys. J., {\bf 187}, 425 (1974)

\bibitem{Begelman:80}
	M. Begelman, R. Blandford, and M. Rees,
	Nature {\bf 287}, 307 (1980).

\bibitem{Merritt:01}
	D. Merritt and F. Cruz,
	Astrophys. J. {\bf 551}, L41 (2001).

\bibitem{Milos:01}
	M. Milosavljevi\'c and D. Merritt,
	Astrophys. J., {\bf 563}, 34 (2001).

\bibitem {Peters:64}
	P. C. Peters,
	Phys. Rev. B {\bf 136}, 1224 (1964).

\bibitem{Berezinsky:97}
	V. Berezinsky, M. Kachelriess, and A. Vilenkin,
	Phys. Rev. Letters, {\bf 79}, 4302 (1997).

\bibitem{Somerville:99}
	R. S. Somerville and T. S. Kolatt,
	Mon. Not. Royal Astron. Soc. {\bf 305}, 1 (1999).

\bibitem{Somerville:00}
	R. S. Somerville, G. Lemson, T. S. Kolatt, and A. Dekel,
	Mon. Not. Royal Astron. Soc. {\bf 316}, 479 (2000).

\bibitem{Wechsler:02}
	R. H. Wechsler, J. S. Bullock, J. R. Primack, A. V.
	Kravtsov, and A. Dekel, 
	Astrophys. J., in press.

\bibitem{Moore:99}
	B. Moore, S. Ghigna, F. Governato, G. Lake, T. Quinn,
	J. Stadel, and P. Tozzi, 
	Astrophys. J. Lett. {\bf 524}, L19 (1999).

\bibitem{Navarro:97}
	J. F. Navarro, C. S. Frenk, and S. D. M. White, 
	Astrophys. J., {\bf 490}, 493 (1997).

\bibitem{Klypin:01a}
	A. A. Klypin, A. V. Kravtsov, J. S. Bullock, and J. R. Primack,
	Astrophys. J., {\bf 554}, 903 (2001).

\bibitem{Merritt:85}
	D. Merritt,
	Astron. J. {\bf 90}, 1027 (1985).

\bibitem{Merritt:98}
	D. Merritt and G. Quinlan,
	Astrophys. J. {\bf 498}, 625 (1998).

\bibitem{Klypin:01b}
	A. Klypin, H.-S. Zhao, and R. S. Somerville,
	astro-ph/0110390

\bibitem{Dorband:01}
	E. N. Dorband, M. Hemsendorf, and D. Merritt, 
	astro-ph/0112092 (2001).

\bibitem{Dehnen:98}
	W. Dehnen and J. Binney,
	Mon. Not. Royal Astron. Soc. {\bf 294}, 429 (1998).

\bibitem{Milos:02a}
	M. Milosavljevi\'c, in preparation (2002).

\bibitem{Barnes:99}
	J. E. Barnes,
	in Proceedings of IAU Symposium 186, Galaxy Interactions
	at Low and High Redshift, ed. J. E. Barnes and D. B. Sanders,
	p. 137 (1999). 

\bibitem{Quinlan:96}
	G. D. Quinlan,
	New Astronomy, {\bf 1}, 35 (1996).

\bibitem{Ullio:01a}
	P. Ullio,
	hep-ph/0105052 (2001).

\bibitem{Milos:02b}
	M. Milosavljevi\'c, D. Merritt, A. Rest, and F. van den Bosch,
	astro-ph/0110185.

\bibitem{Bekenstein:73}
	J. Bekenstein,
	Astrophys. J. {\bf 183}, 675 (1973).

\bibitem{Fitchett:83}
	M. Fitchett, 
        Month. Not. Royal Astronom. Soc. {\bf 203}, 1049 (1983). 

\bibitem{Eardley:83} 
	D. M. Eardley, in Gravitational Radiation, 
	edited by N. Deruelle and T. Piran
	(North-Holland, Amsterdam, 1983), p. 257.

\bibitem{Redmount:89}
	I. H. Redmount and M. J. Rees, 
	Comm. Astrophys. {\bf 14}, 165 (1989).

\bibitem{Nakano:99}
	T. Nakano and J. Makino,
	Astrophys. J. {\bf 525}, 77 (1999)

\bibitem{Ullio:01b}
	P. Ullio, H.-Z. Zhao, and M. Kamionkowski,
	Phys. Rev. D., {\bf 64}, 043504 (2001).

\end{thebibliography}
\end{document}